\documentclass[sigconf,screen]{acmart}
\AtBeginDocument{%
  \providecommand\BibTeX{{%
    \normalfont B\kern-0.5em{\scshape i\kern-0.25em b}\kern-0.8em\TeX}}}

\copyrightyear{2024}
\acmYear{2024}
\setcopyright{acmlicensed}
\acmConference[ICSE '24]{2024 IEEE/ACM 46th International Conference on Software Engineering}{April 14--20, 2024}{Lisbon, Portugal}
\acmBooktitle{2024 IEEE/ACM 46th International Conference on Software Engineering (ICSE '24), April 14--20, 2024, Lisbon, Portugal} 
\acmDOI{10.1145/3597503.3639153} 
\acmISBN{979-8-4007-0217-4/24/04}

\acmConference[ICSE 2024]{46th International Conference on Software Engineering}{April 2024}{Lisbon, Portugal}
\usepackage{makecell}

\usepackage{algorithmic}
\usepackage{graphicx}
\usepackage{xcolor}
\usepackage{tabularx}

\usepackage{multirow}
\usepackage{enumerate}
\usepackage{listings, xcolor}
\definecolor{verylightgray}{rgb}{.97,.97,.97}
\lstdefinelanguage{Solidity}{
  keywords=[1]{anonymous, assembly, assert, balance, break, call, callcode, case, catch, class, constant, continue, constructor, contract, debugger, default, delegatecall, delete, do, else, emit, event, experimental, export, external, false, finally, for, function, gas, if, implements, import, in, indexed, instanceof, interface, internal, is, length, library, log0, log1, log2, log3, log4, memory, modifier, new, payable, pragma, private, protected, public, pure, push, require, return, returns, revert, selfdestruct, send, solidity, storage, struct, suicide, super, switch, then, this, throw, transfer, true, try, typeof, using, value, view, while, with, addmod, ecrecover, keccak256, mulmod, ripemd160, sha256, sha3}, 
  keywordstyle=[1]\color{blue}\bfseries,
  keywords=[2]{address, bool, byte, bytes, bytes1, bytes2, bytes3, bytes4, bytes5, bytes6, bytes7, bytes8, bytes9, bytes10, bytes11, bytes12, bytes13, bytes14, bytes15, bytes16, bytes17, bytes18, bytes19, bytes20, bytes21, bytes22, bytes23, bytes24, bytes25, bytes26, bytes27, bytes28, bytes29, bytes30, bytes31, bytes32, enum, int, int8, int16, int24, int32, int40, int48, int56, int64, int72, int80, int88, int96, int104, int112, int120, int128, int136, int144, int152, int160, int168, int176, int184, int192, int200, int208, int216, int224, int232, int240, int248, int256, mapping, string, uint, uint8, uint16, uint24, uint32, uint40, uint48, uint56, uint64, uint72, uint80, uint88, uint96, uint104, uint112, uint120, uint128, uint136, uint144, uint152, uint160, uint168, uint176, uint184, uint192, uint200, uint208, uint216, uint224, uint232, uint240, uint248, uint256, var, void, ether, finney, szabo, wei, days, hours, minutes, seconds, weeks, years},  
  keywordstyle=[2]\color{teal}\bfseries,
  keywords=[3]{block, blockhash, coinbase, difficulty, gaslimit, number, timestamp, msg, data, gas, sender, sig, value, now, tx, gasprice, origin},  
  keywordstyle=[3]\color{violet}\bfseries,
  identifierstyle=\color{black},
  sensitive=false,
  comment=[l]{//},
  morecomment=[s]{/*}{*/},
  commentstyle=\color{gray}\ttfamily,
  stringstyle=\color{red}\ttfamily,
  morestring=[b]',
  morestring=[b]"
}
\usepackage{colortbl}

\usepackage{tikz,pifont}
 
\newcommand{\para}[1]{\vspace{2pt}\noindent\textbf{#1.~}}

\usepackage{threeparttable}
\usepackage{wasysym}
\begin{document}

\title{Uncover the Premeditated Attacks: Detecting Exploitable Reentrancy Vulnerabilities by Identifying Attacker Contracts}

\author{Shuo Yang}
\affiliation{%
  \institution{Sun Yat-sen University}
  \city{Zhuhai}
  \country{China}
}
\email{yangsh233@mail2.sysu.edu.cn}

\author{Jiachi Chen}
\authornote{corresponding author}
\affiliation{%
  \institution{Sun Yat-sen University}
  \city{Zhuhai}
  \country{China}
}
\email{chenjch86@mail.sysu.edu.cn}

\author{Mingyuan Huang}
\affiliation{%
  \institution{Sun Yat-sen University}
  \city{Zhuhai}
  \country{China}
}
\email{huangmy83@mail2.sysu.edu.cn}

\author{Zibin Zheng}
\affiliation{%
  \institution{Sun Yat-sen University}
  \city{Zhuhai}
  \country{China}}
\email{zhzibin@mail.sysu.edu.cn}

\author{Yuan Huang}
\affiliation{%
  \institution{Sun Yat-sen University}
  \city{Zhuhai}
  \country{China}
}
\email{huangyuan5@mail.sysu.edu.cn}

\begin{abstract}
Reentrancy, a notorious vulnerability in smart contracts, has led to millions of dollars in financial loss. However, current smart contract vulnerability detection tools suffer from a high false positive rate in identifying contracts with reentrancy vulnerabilities. Moreover, only a small portion of the detected reentrant contracts can actually be exploited by hackers, making these tools less effective in securing the Ethereum ecosystem in practice. 

In this paper, we propose BlockWatchdog, a tool that focuses on detecting reentrancy vulnerabilities by identifying attacker contracts. These attacker contracts are deployed by hackers to exploit vulnerable contracts automatically. By focusing on attacker contracts, BlockWatchdog effectively detects truly exploitable reentrancy vulnerabilities by identifying reentrant call flow. 
Additionally, BlockWatchdog is capable of detecting new types of reentrancy vulnerabilities caused by poor designs when using ERC tokens or user-defined interfaces, which cannot be detected by current rule-based tools.
We implement BlockWatchdog using cross-contract static dataflow techniques based on attack logic obtained from an empirical study that analyzes attacker contracts from 281 attack incidents. 
BlockWatchdog is evaluated on 421,889 Ethereum contract bytecodes and identifies 113 attacker contracts that target 159 victim contracts, leading to the theft of Ether and tokens valued at approximately 908.6 million USD. Notably, only 18 of the identified 159 victim contracts can be reported by current reentrancy detection tools.
\end{abstract}





\maketitle

\section{introduction}

In recent years, Ethereum has experienced significant growth in popularity and market cap~\cite{etherscan}, primarily due to its ability to support a wide range of decentralized applications (DApps)~\cite{wu2019empirical}, such as decentralized finance (DeFi)~\cite{werner2022sok} and non-fungible tokens (NFTs)~\cite{nftdefects}. This is made possible through the use of smart contracts~\cite{zheng2020overview}, which are Turing-complete programs that run on the blockchain.
However, as the value of Ethereum continues to rise, attackers are increasingly targeting contracts with vulnerabilities that can be exploited to make unfair profits. Reentrancy is one of the infamous vulnerabilities, which has caused huge financial losses~\cite{zheng2023turn} since the 150 million USD DAO attack in 2016~\cite{li2020survey}, and new reentrancy attacks keep popping up in more and more complex forms.
For example, an attacker exploited a reentrancy vulnerability to drain approximately 1,300 ETH (1.43 million USD) from the NFT money market platform called Omni~\cite{omni_hacker} by using the hook function \textit{onERC721Received()} declared in the ERC721 standard~\cite{eip721}.

Numerous studies have focused on detecting vulnerabilities in smart contracts~\cite{luu2016making,grech2018madmax,krupp2018teether}, proposing various methods such as static analysis and dynamic testing to identify potential issues. However, these works face two main limitations.
First, there is a high false-positive rate in detecting reentrancy vulnerabilities~\cite{zheng2023turn}, as they cannot correctly detect some protection patterns.
Furthermore, these methods mainly focus on reentrancy caused by \textit{call.value()} operations, which cannot cover more complex reentrancy vulnerabilities (leading to false negatives) caused by poor designs when using standard ERC tokens, e.g., ERC721,
or user-defined interfaces (see Section~\ref{sec:newreen}).
Second, only 2.68\% of contracts with reentrancy vulnerabilities can truly be exploited by hackers, and only 0.4\% of the Ethers at stake could be exploited~\cite{perez2021smart}.
Real-world attackers intend to evaluate the cost and benefit of an attack, but most contracts do not have the balance that can be extracted. Specifically, on Ethereum, only 3\% of the contracts have a non-zero balance. Thus, most of the vulnerable contracts labeled by the tools cannot be exploited and are false alarms, which makes it less effective in securing the Ethereum ecosystem in practice.

Exploiting reentrancy vulnerabilities requires deploying malicious contracts that initiate callbacks to the victim contract. In this paper, we shift our focus from vulnerability detection to analyzing attacker contracts. 
To investigate how attackers implement callback logic on victims, we conduct an empirical study by analyzing 281 attack incident reports from various platforms, e.g., Twitter~\cite{twitter}, Medium~\cite{medium}, and Peckshield~\cite{peckshield}, spanning from June 2016 to July 2022. 
These platforms provide comprehensive and timely descriptions of attack incidents, which are also adopted by other works~\cite{nftdefects}.
Consequently, we summarize three types of reentrancy attack types based on the functions that attacker contracts used to make callbacks (see Section~\ref{sec:reen}).
Furthermore, we propose BlockWatchdog, a tool that utilizes cross-contract static dataflow techniques to identify reentrancy attacker contracts. \textit{First}, BlockWatchdog decompiles the contract's bytecode to the intermediate representation (IR) and extracts flow and external call information in the functions. \textit{Second}, BlockWatchdog identifies the contracts in the call chain and constructs the cross-contract control flow graph (xCFG) and the cross-contract call graph (xCG) of the contract based on dataflow rules. \textit{Then}, it traces all call chains to perform a taint analysis to determine whether the attacker can manipulate the call chain, making itself called again to implement reentrancy.
Based on detection patterns designed in collaboration with external call and dataflow information, BlockWatchdog reports whether the contract is an attacker contract or not and identifies vulnerable victim contracts.

In the experiment, we first evaluate BlockWatchdog on our collected ground-truth dataset, which contains 18 reentrancy attacker contracts. Then, we run BlockWatchdog on a real-world dataset containing 421,889 contracts' bytecodes obtained by replaying transactions from block number 10 million to 15.5 million on the Ethereum mainnet. The average detection time of it is 17.66 seconds. Furthermore, BlockWatchdog identifies 249 attacker smart contracts in this dataset, and 113 of them are labeled as true positives. Among them, 40 are 0-day attacker contracts, which involve 159 victim contracts. Ethers and tokens worth approximately 908.6 million USD in these contracts have been stolen by attackers. Furthermore, we run seven tools for reentrancy vulnerability detection on identified victim contracts; only 18 (11.3\%) of them can be correctly reported.

The main contributions of our work are as follows.

\begin{itemize}
    \item We shift the detection focus from vulnerable contracts to attacker contracts, which alleviates the high false positive problem and limited capability of current tools in finding reentrancy.
    \item We summarize three types of reentrancy attacks from an empirical study and introduce BlockWatchdog, a cross-contract static dataflow analysis tool to find attacker contracts and vulnerable victim contracts they target. Additionally, BlockWatchdog is extensible for users to program more rules to cover new attack types.
    \item We evaluate the performance of BlockWatchdog on a dataset consisting of 421,889 contracts bytecode. Our experiments show that BlockWatchdog identifies 113 attacker contracts and 159 victim contracts, which hold Ethers and tokens worth approximately 908.6 million USD. Only 18 of the 159 victims can be detected by the current tools. We publicize the source code of BlockWatchdog and the experimental results in our repository\footnote{\href{https://github.com/shuo-young/BlockWatchdog}{https://github.com/shuo-young/BlockWatchdog}}.
\end{itemize}


\section{Background and Motivation}\label{sec:background}
\subsection{Solidity Smart Contracts}
A smart contract is a self-executing agreement that is enforced by the rules encoded in its code~\cite{szabo1997formalizing}. 
Solidity is the most popular programming language for smart contracts on Ethereum. The bytecode and transactions of the deployed smart contracts are permanently stored on the blockchain~\cite{zheng2020overview}. The immutability of smart contracts ensures that their code and behavior cannot be modified once deployed, and they execute automatically based on their predefined logic.
Ethereum Virtual Machine (EVM) is a stack-based virtual machine that executes transactions by splitting the EVM bytecode into operation codes (opcodes) and following their instructions. 

\subsection{Reentrancy}\label{sec:newreen}
The reentrancy vulnerability has resulted in significant financial losses over the past few years. 
There are many works that focus on detecting reentrancies caused by \textit{call.value()} pattern~\cite{zheng2023turn}.
Solidity smart contracts have a unique mechanism that requires any contract that receives Ethers to implement a fallback function. The fallback function will be executed if the contract receives Ether from other addresses. 
If the victim contract transfers Ethers to the malicious attacker contract, the malicious one can take over the control flow and repetitively call the victim in its fallback function. Many attackers have exploited this fallback mechanism to drain funds from victims.
Not only those caused by \textit{call.value()}, there are some new reentrancy types. For example, Lenf.me~\cite{lendfme} and Omni~\cite{omni_hacker} were attacked by the bad design of using ERC777~\cite{erc777} and ERC721~\cite{eip721} tokens, respectively. 

In addition, poor design when using user-defined interfaces can also lead to reentrancy issues.
Figure~\ref{fig:ivisor} shows the attacker contract that hacked 8.2 million USD through a reentrancy attack on IVisor~\cite{ivisor}, a liquidity management protocol of Uniswap V3~\cite{uniswap_v3}. The function \textit{delegatedTransferERC20()} (L25) is defined by the developers, which is not declared in the ERC token standard.
The attacker contract injects external calls into the function \textit{delegatedTransferERC20()} (L13-L17) invoked by the victim contract \textit{RewardHypervisor} (L20).
In detail, the attacker contract invokes the function \textit{deposit()} (L8) of the contract \textit{RewardHypervisor}. Then, \textit{RewardHypervisor} calls the \textit{delegatedTransferERC20()} (L25) of contract \textit{from}, which is set by the attacker contract with \textit{address(this)}, i.e., the attacker contract itself (L25). However, the attacker makes a callback to \textit{RewardHypervisor} again to deposit again on line 8, which makes it suffer from a reentrancy vulnerability. The contract \textit{RewardsHypervisor} does not contain \textit{call.value()} reentrancy vulnerability type, which current detection tools focus on. Yet, it was still attacked by the malicious attacker contract to make unfair gains due to the bad design when using user-defined interface \textit{delegatedTransferERC20()}. 

\begin{figure}[!htbp]
\begin{lstlisting}
 // the decompiled IR of the attacker contract bytecode
 contract Attacker {
   function 0x4a0b0c38() public payable { 
     0x28e();}

   function 0x28e() private {
     require(_pool.code.size);
     v0, v1 = _pool.deposit(0x52b7d2dcc80cd2e4000000, address(this), _admin);
     require(v0);
     require(RETURNDATASIZE() >= 32);
     return ;}

   function delegatedTransferERC20(address varg0, address varg1, uint256 varg2) public payable { 
     require(msg.data.length - 4 >= 96);
     _count += 1;
     if (_count < 2) {
       0x28e();}}
 }
 // the source code of the victim contract
 contract RewardsHypervisor {
   function deposit(uint256 visrDeposit, address payable from, address to) external returns (uint256 shares) {
   ...
   if(isContract(from)) {
     require(IVisor(from).owner() == msg.sender); 
     IVisor(from).delegatedTransferERC20(address(visr), address(this), visrDeposit);}
   else {
     visr.safeTransferFrom(from, address(this), visrDeposit);}
     vvisr.mint(to, shares);}
 }
\end{lstlisting}
\caption{Reentrancy attack toward Visor Finance.}
\label{fig:ivisor}
\end{figure}

\subsection{Prior Research and Their Limitations}
\para{Unexploitable for the detected contracts}
Previous research has focused on detecting reentrancy vulnerabilities, but most of the contracts detected are either toy contracts with no value or cannot be exploited by attackers~\cite{perez2021smart}. Specifically, only 1.98\% of the 23,327 reported vulnerable contracts from six academic projects have been exploited since deployment, affecting only 0.27\% of the funds in the contracts~\cite{luu2016making,tsankov2018securify,kalra2018zeus,grech2018madmax,nikolic2018finding,krupp2018teether}. The reason is that the majority of Ether or tokens are held by only a small number of vulnerable contracts, which are lucrative for hackers, making most of the detected contracts unexploitable. However, detecting attacker contracts can help identify truly exploitable yet vulnerable contracts.

\para{Poor performance in detecting reentrancy}\label{sec:userdefine}
Existing reentrancy detection tools have an extremely high false positive rate. More than 99.8\% of the reentrant contracts detected by the tools~\cite{mueller2018smashing,luu2016making,choi2021smartian,rao2012sailfish,tsankov2018securify} are false positives~\cite{zheng2023turn}, as these tools do not detect some protection patterns, such as the reentrancy lock. To reduce false positives, detecting attacker contracts can help find those hackers who aim at truly exploitable contracts without reentrancy protections. Furthermore, existing tools cannot cover new types of reentrancies caused by poor design when using ERC tokens, which can result in false negatives. This is because rule-based tools have limited scalability, as each rule can only check a specific reentrancy type (e.g., reentrancy caused by \textit{call.value()}) and cannot cover emerging types, such as new ERC standards~\cite{norvill2019standardising} (e.g., reentrancy caused by ERC721 and ERC1155~\cite{erc1155}), or user-defined interfaces, e.g., \textit{delegateTransferERC20()} in Figure~\ref{fig:ivisor}.

To address these limitations, a more general detection method is needed to reduce both false positives and false negatives.
Reentrancy vulnerabilities are mainly characterized by the mutual invocation of the victim contract and the attacker contract with callback flow features, but victim contracts contain limited and non-homogeneous information, making it challenging to summarize a generic signature for rule-based tools. It motivates us to recover reentrancy features from the attacker contracts perspective.


\section{Attacker Smart Contracts for Reentrancy}\label{sec:reen}
\begin{figure}[h]
	\centering
	\includegraphics[width=\linewidth]{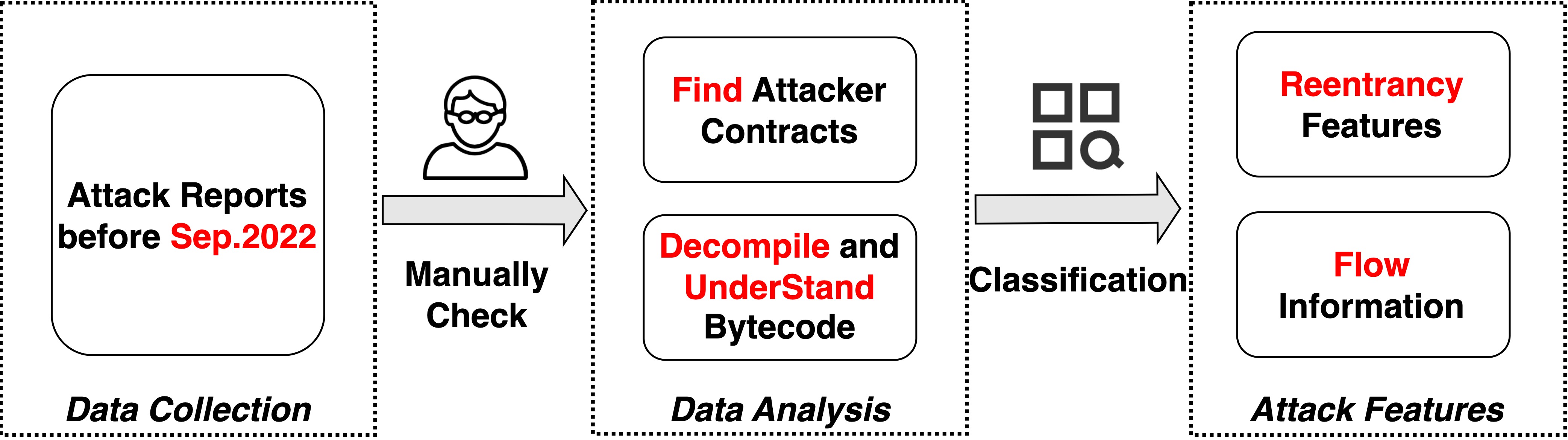}
	\caption{Workflow of finding new reentrancy types.} \label{fig:data_analysis}
\end{figure}

In this section, we present an empirical study aimed at identifying the characteristics of attacker contracts involved in historical reentrancy attacks. Our goal is to uncover reentrancy vulnerabilities that existing tools have missed. We approach the analysis of attacker contracts through a data collection process, followed by a data analysis and feature identification process, as illustrated in Figure~\ref{fig:data_analysis} and described in subsequent subsections.

\subsection{Data Collection}
To gain a comprehensive understanding of attacks, it is necessary to collect and analyze relevant data. We began by searching for attack reports published by reputable blockchain security firms such as PeckShield, as well as information from social media platforms like Twitter and publishing platforms like Medium. In total, we identified 281 incidents that occurred between June 2016 and July 2022. For each report, we recorded key information such as the target project, the victim contract address, the attack time, and the associated losses. All the data collected and their source links are available for public access in our open repository.

\subsection{Data Analysis}\label{sec:attackercontracts}
\para{Attacker contract identification}
As we do not know how hackers perform the attack, we intend to find the attacker contracts to analyze the attack logic from the collected data. Specifically, two of our authors, both with more than two years of experience in blockchain security, manually analyzed the 281 incidents collected using the open card sorting approach~\cite{spencer2009card}. During the manual check, we identified two distinct types of attacks: those directly attacked by externally owned accounts (EOAs) of hackers and those attacked by attacker contracts that hackers deployed.

\begin{table}[htbp]
\caption{Attack Types Obtained from Collected Data}\label{tab:taxonomy}
\centering
    \scalebox{0.94}{
        \begin{tabular}{c||c|c|c|c|c|c|c|c|c}
            \hline
              & DoS & BR & IO & RE & IA & CI & CAD & FL & Others \\ \hline
            AC/EOA & \LEFTcircle              & \CIRCLE        & \Circle                   & \CIRCLE              & \Circle             & \Circle     & \Circle & \CIRCLE & \LEFTcircle             \\ \hline
            CA & \frownie                & \frownie      & \frownie                & \frownie            & \frownie               & \frownie        & \frownie  & \frownie  & \frownie          \\ \hline
        \end{tabular}\label{fig:alltypes}
    }
\end{table}

We then classify the attack types based on the necessity of an attacker contract and the availability of the attacker contract's source code. Table~\ref{tab:taxonomy} shows the eight types of attack that we identified from our 281 collected incidents. We use the ``Others'' category to cover attacks that target specific design flaws of victims, e.g., infinite approval to vulnerable contracts~\cite{primitivefinance}. The $\CIRCLE$ and $\Circle$ symbols in the column ``Attacker Contract'' represent attack types that require deploying the attacker contract (AC) or using EOA transactions, respectively. The symbol $\LEFTcircle$  represents attack types that do not require the deployment of an attacker contract in some cases. The $\frownie$ symbol in the ``Code Availability'' (CA) column represents attacker contracts whose source code is not available. Among the 281 samples, we found 31 attacker contracts from 28 reports, classified as Denial of Service (DoS), Bad Randomness (BR), Reentrancy (RE), Flashloan (FL), and Others. Attacks such as Integer Overflow (IO), Improper Authentication (IA), Call Injection (CI), and Call-after-destruct (CAD), which do not involve attacker contracts, are out of the scope of our analysis. As this paper focuses on new types of reentrancy vulnerabilities missed by existing vulnerability detection tools, we will illustrate reentrancy vulnerabilities from the perspective of the 18 reentrancy attacker contracts collected.

\begin{figure}[h]
	\centering
	\includegraphics[width=\linewidth]{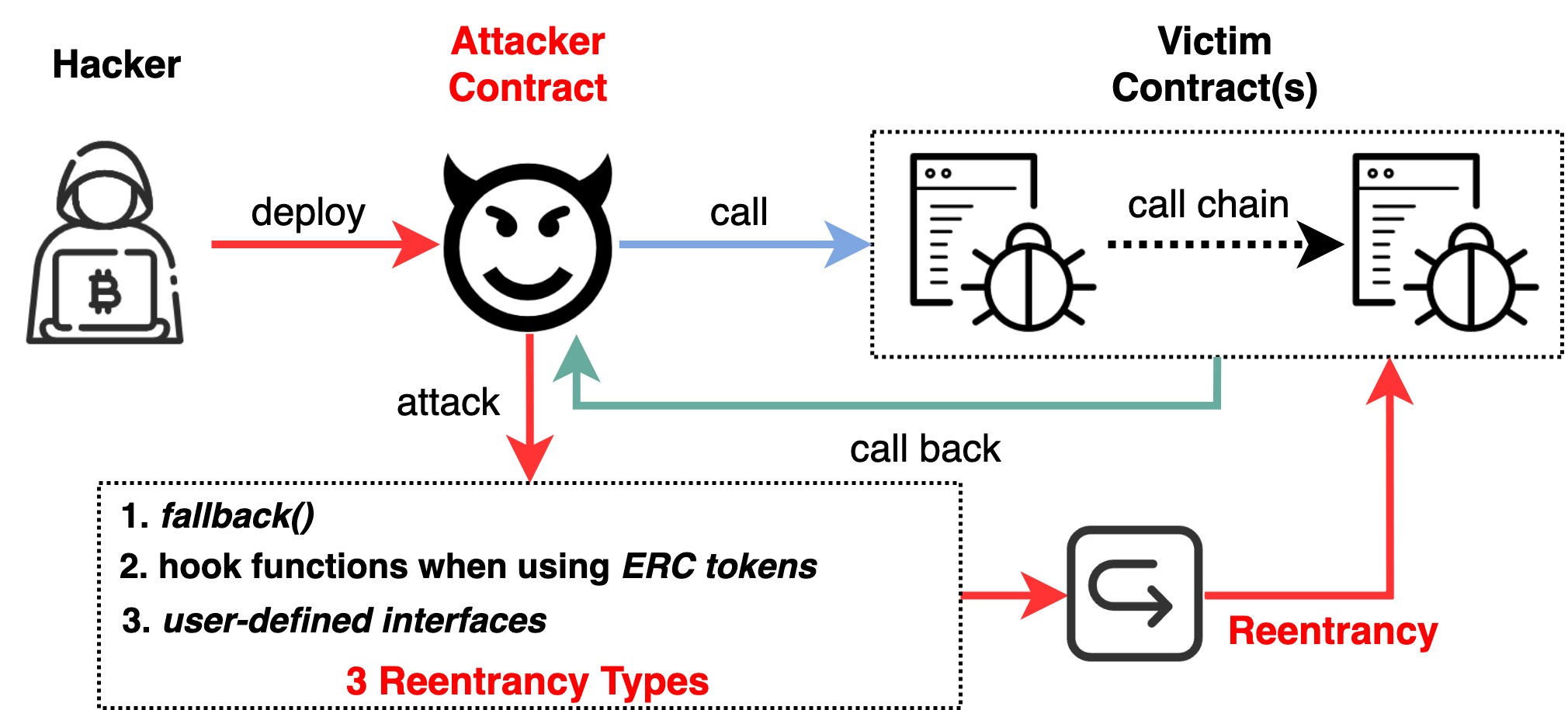}
	\caption{An overview of reentrancy attacks.} \label{fig:attack_overview}
\end{figure}

\para{Decompilation and understanding}  
To gain a deeper understanding of the attacker behaviors employed by reentrancy attacker contracts, we collected the bytecode of attacker contracts for analysis, as none of the identified attacker contracts released their source code to the public. We decompiled the EVM bytecode to recover a readable intermediate representation (IR) of the attacker contract. We then followed the attack process and description according to the report to understand how the attacker implements the attack logic from its bytecode. Figure~\ref{fig:attack_overview} provides an overview of how attacker contracts exploit victim contracts. The hacker first programs the attacker contract and deploys it on the blockchain. Subsequently, the attacker contract automatically executes and initiates external calls to the victim contracts. Notably, in reentrancy attacks, the attacker contract can pass parameters to victim contracts, which makes them call back to the attacker contract again, and there can be multiple victims in this call chain.
As shown in Figure~\ref{fig:attack_overview}, we summarize three types of reentrancy based on the functions utilized by attacker contracts to perform reentrancy, i.e., (1) \textit{fallback()}, poor designs when using (2) \textit{ERC tokens}, or (3) \textit{user-defined interfaces}.
Specifically, the attacker contract can implement reentrancy logic in the \textit{fallback()} function when receiving Ethers. It can also inject callbacks into hook functions when using ERC tokens, e.g., hook function \textit{onERC721Received} when using ERC721, or user-defined interfaces, e.g., the case shown in Figure~\ref{fig:ivisor}, to implement reentrancy.


\subsection{Attacker Contract Features}
\begin{figure}[h]
	\centering
	\includegraphics[width=\linewidth]{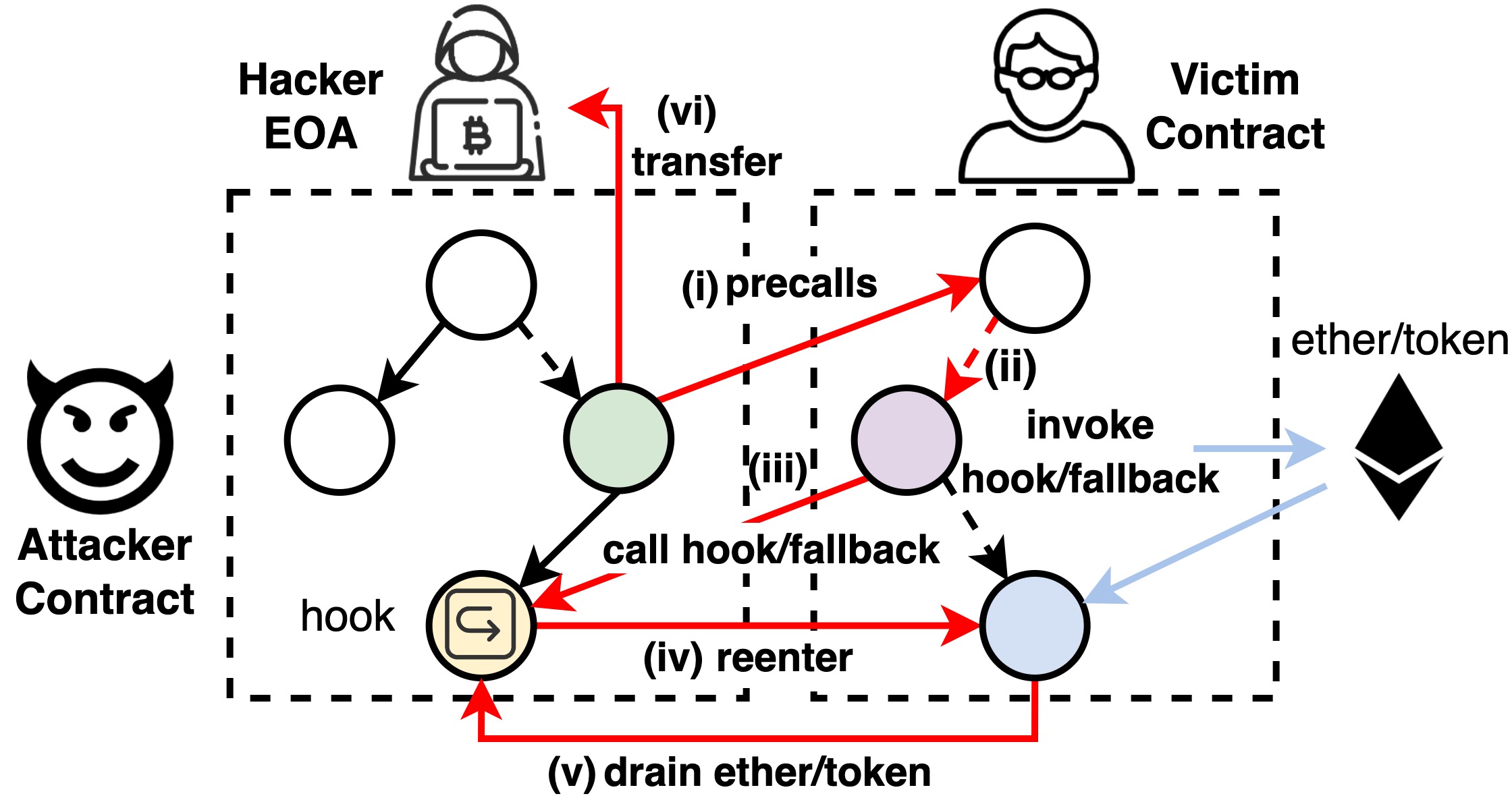}
	\caption{An example to illustrate reentrancy attack.} \label{fig:reentrancy}
\end{figure}

Figure~\ref{fig:reentrancy} shows an example that illustrates the high-level features of the reentrancy attack focusing on the call flow.
To perform the reentrancy attack, the attacker contract (1) first calls the victim contract (step i in Figure~\ref{fig:reentrancy}) to make the victim invoke a callback (step ii) to the attacker's hook function or transfer Ether to the attacker contract (step iii), (2) then the attacker contract calls the victim again in the hook function or the fallback function, and reenters (step iv) to invoke functions that can generate unfair profits (step v), (3) next, the profits can be transferred to the attacker EOA (step vi) to complete the reentrancy attack.
This call flow information shows how the attacker contract interacts with the victim contract.
The summarized three types of reentrancy help us identify the reentrancy from function-level call information.
The design of our method shown in the following section is based on these features obtained from our empirical study.


\section{Methodology}\label{sec:method}
In this section, we introduce the BlockWatchdog tool, which can detect attacker contracts that perform reentrancy attacks. We first give an overview of the approach and then provide the details from the perspectives of flow information extraction, cross-contract static analysis, and attacker contract detection.

\subsection{Overview}
\begin{figure}[htbp]
    \centering
    \includegraphics[width=\linewidth]{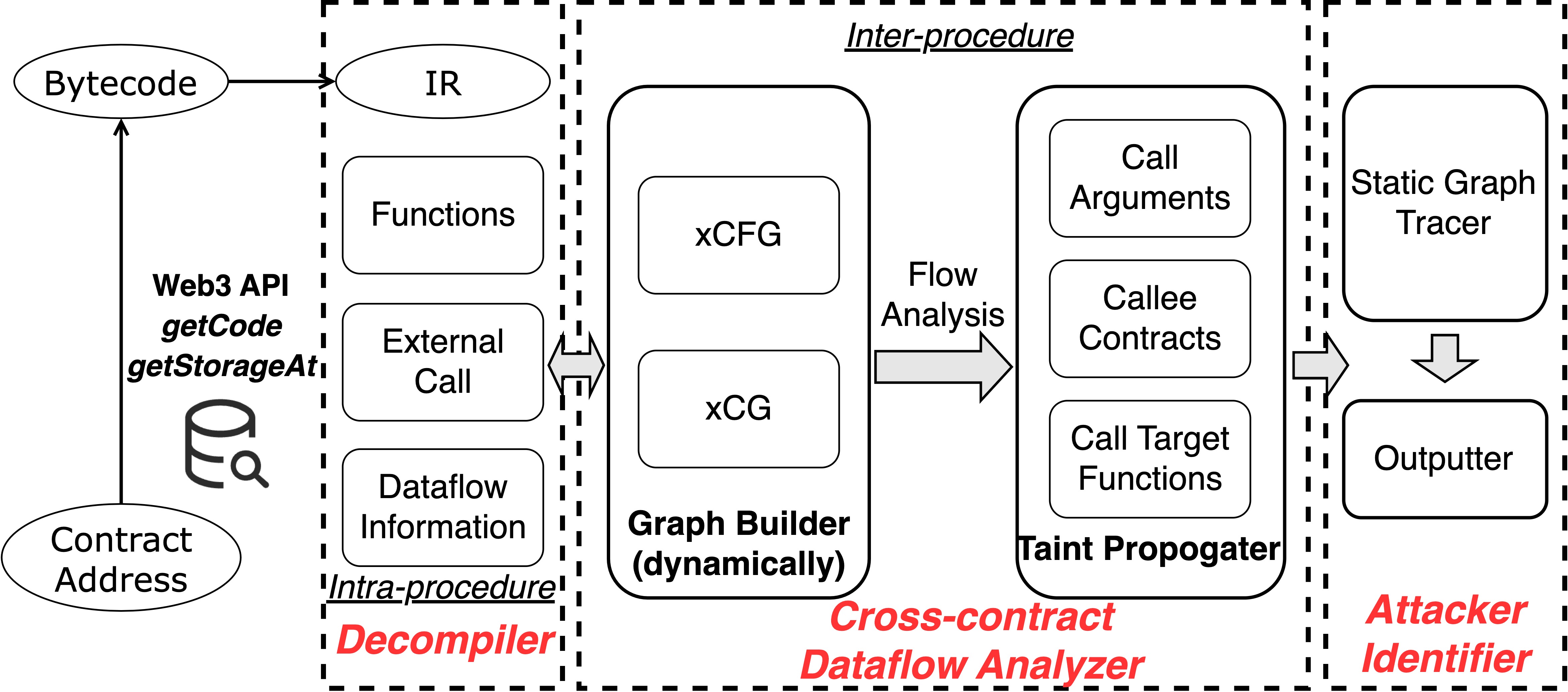}
    \caption{An overview of the approach of BlockWatchdog.}
    \label{fig:overview}
\end{figure}

BlockWatchdog consists of three main components: \textit{Decompiler}, \textit{Cross-Contract Dataflow Analyzer}, and \textit{Attack Identifier}. 
Figure~\ref{fig:overview} shows an overview of the BlockWatchdog approach.
The tool can accept a contract bytecode or a real address on Ethereum as input. If a contract address is provided, the tool retrieves the bytecode from the Web3 API~\cite{web3py}. BlockWatchdog decompiles the bytecode to the IR and extracts critical flow information for dataflow analysis in \textit{Decompiler}. Next, the \textit{Cross-contract Dataflow Analyzer} constructs the xCFG and xCG of the contract according to the information obtained from the decompilation. We use xCFG and xCG to bridge the flow information of all contracts in the call chains for taint analysis. Taints originating from the attacker contract are propagated through the function call arguments and returns based on designed transfer rules while tracing all possible call chains in the xCG. Finally, the \textit{Attack Identifier} identifies and reports the three attack types based on our detection patterns based on the result of the taint flow analysis. 

To implement BlockWatchdog, we adopt a public node provided by Alchemy~\cite{alchemy} to request data from the blockchain. Specifically, we use the Web3 API \textit{getCode} to fetch the bytecode of a contract and use \textit{getStorageAt} to obtain the storage data in a specific slot and offset from a contract account. For decompilation, we use the EVM bytecode decompiler Elipmoc, which improves over all the notable past decompilers~\cite{grech2022elipmoc,kong2023defitainter}. Elipmoc can disassemble the EVM bytecode into EVM opcodes and construct the control flow graph based on identifying flow-related opcodes like JUMP and JUMPI. The function borders and the IR are then recovered for further analysis.

\subsection{Flow Information Extraction}
When analyzing an input contract, the first step in BlockWatchdog is to decompile its EVM bytecode to the IR. In the \textit{Decompiler} component, we extract flow information from the IR to enable dataflow analysis and attacker vector identification.

\para{Constant contract address and function signature identification} 
To conduct cross-contract static dataflow analysis, we need to identify the target contracts and functions that the input contract intends to call. Based on our analysis of previous attack incidents, we observe that attacker contracts typically hard-code the contract address with which they want to interact or store the target contract's address in storage. Therefore, we use the decompiler to identify the constant value (conforming to EIP-55~\cite{eip55}) of the callee contract address. The slot number with specific offsets to locate the storage address can be obtained through the Web3 API \textit{ getStorageAt}. Additionally, we identify the function signatures through decompilation, which specifies which functions can be invoked in each external call. For each external call identified in the caller contract, we obtain the callee contract and function signature to locate the callee site of the contract being called, allowing us to construct the entire xCFG and xCG for inter-procedure cross-contract analysis.

\begin{figure}[htbp]
\centering
    \includegraphics[width=\linewidth]{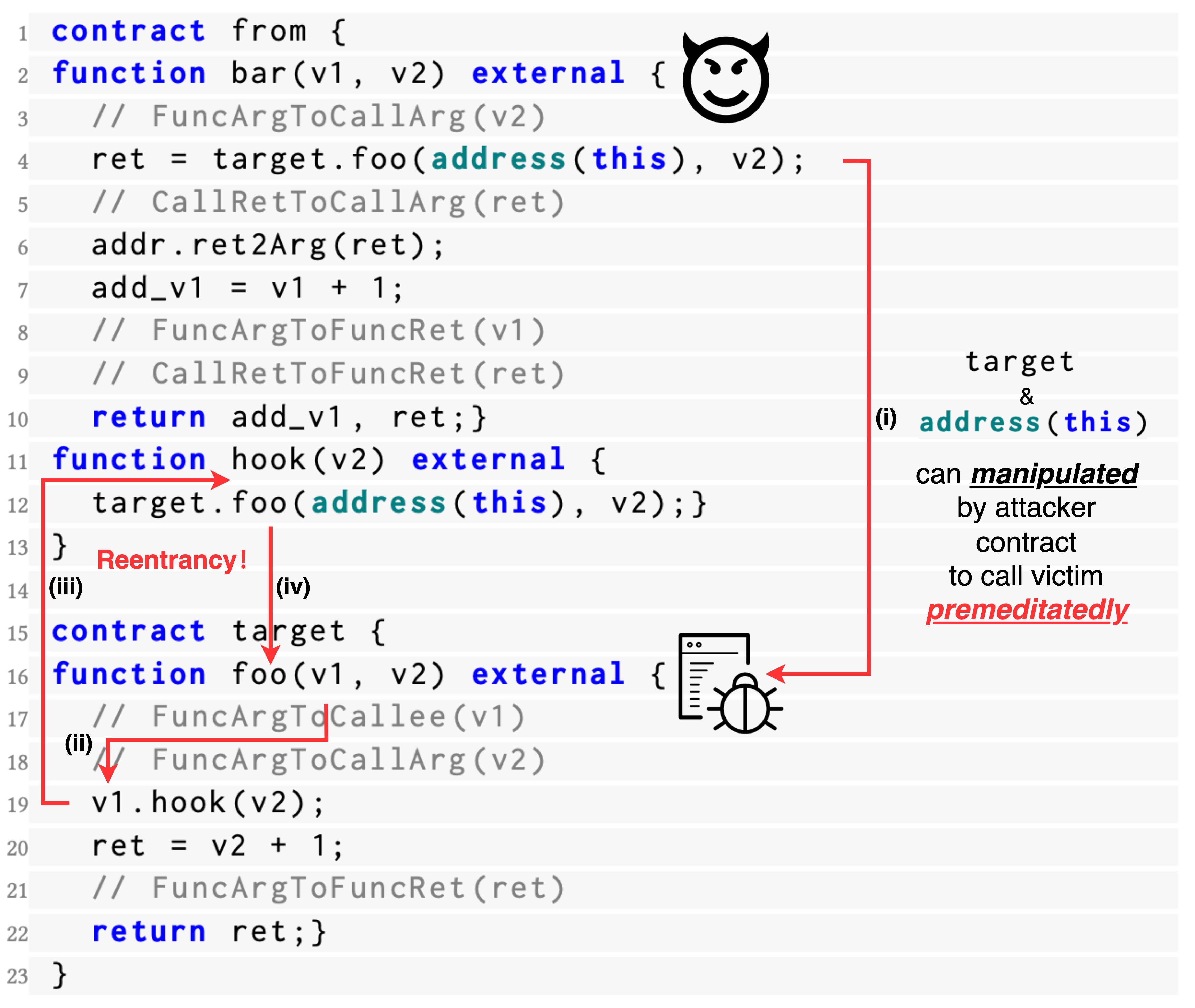}
\caption{Toy contracts for flow information illustration.}
\label{fig:flow}
\end{figure}

\para{Flow and external call information extraction}
To extract the flow information from a contract, we first obtain every entrance and exit of the contract. For every public function identified by the decompiler, we extract its function arguments and return parameters, which denote the start and end points of the dataflow within the function, respectively. Function arguments can also flow to the parameters of external calls made within the function, while the return values of external calls can flow back to the function's return parameters. Therefore, we design five dataflow rules in Table~\ref{tab:flow} for each public function in the contract. We illustrate each rule based on Figure~\ref{fig:flow}.
Taking the \textit{FuncArgToCallArg} and \textit{FuncArgToCallee} as an example, in Figure~\ref{fig:flow}, function argument \textit{v2} of the function \textit{bar()} (L2-L10) flows to the arguments of the external call \textit{target.foo()} (L4) by rule \textit{FuncArgToCallArg}. In contract \textit{target}, the function argument \textit{v1} of \textit{foo()} (L16-L22), which is set by the caller contract \textit{from} (line (i) in Figure~\ref{fig:flow}), can flow to the callee variable \textit{v1} of the call operation \textit{v1.hook(v2)} (L19) through rule \textit{FuncArgToCallee} (line (ii) in Figure~\ref{fig:flow}). By combining these two rules, we can obtain the flow information that \textit{address(this)} (L4) set by contract \textit{from} can flow to the callee variable \textit{v1} (L19) in contract \textit{target}.
In this case, the contract \textit{from} can make the contract \textit{target} call its implemented function \textit{hook} through the dataflow process. Then, the call flow returns to the contract \textit{from}, which makes it capable of calling back to the \textit{target.foo()} again (L12), and leads to the reentrancy. 
\begin{table}[htbp]
\centering
\caption{Dataflow Rules in Intra-Procedure Analysis}
\resizebox{\columnwidth}{!}{
\begin{tabular}{p{2.2cm}||p{5.6cm}}
\hline
\textbf{Flow Type} & \textbf{Meaning}                    \\ \hline
\textit{FuncArgToCallArg}   & flow from function arguments to call arguments            \\ \hline
\textit{FuncArgToFuncRet}   & flow from function arguments to function returns  \\ \hline
\textit{FuncArgToCallee}    & flow from function arguments to callee variables            \\ \hline
\textit{CallRetToCallArg}   & flow from call returns to call arguments                    \\ \hline
\textit{CallRetToFuncRet}   & flow from call returns to function returns       \\ \hline                                                               
\end{tabular}}
\label{tab:flow}
\end{table}

Additionally, we focus on whether the input contract implements an external call in the hook or fallback function, which attackers can leverage to make other contracts call back and succeed in reentering the attacker contract.
We summarize six hook functions from five Ethereum Improvement Proposals (EIPs)~\cite{eip} that we find to be involved in our collected reentrancy attacker contracts, as shown in Table~\ref{tab:eip}.
Noticeably, we list these hook functions to illustrate the features of the attacker contracts; our BlockWatchdog does not rely on these specific function signatures but focuses on the call flow features.
Furthermore, the fallback function called when receiving Ethers and user-defined interfaces, e.g., \textit{IVisor.delegatedTransferERC20()}, can also be used to perform reentrancy.
Therefore, based on the external call information found in these types of functions and flow information extracted during the intra-procedure analysis, BlockWatchdog determines whether a reentrancy attacker contract is present.

\begin{table}[htbp]
\centering
\caption{Hook Functions Declared by EIP}
\resizebox{\columnwidth}{!}{
\begin{tabular}{l||l|l}
\hline
\textbf{Standard} & \textbf{Function Name}     & \textbf{Function Signature}        \\ \hline
EIP-20            & \textit{transferFrom}          & 0x01c6adc3                                    \\ \hline
EIP-721           & \textit{onERC721Received}     & 0x150b7a02                                        \\ \hline
EIP-1155          & \textit{onERC1155Received}     & 0xf23a6e61                                  \\ \hline
EIP-777           & \textit{tokensToSend}        &  0x75ab9782                               \\ \hline
EIP-777           & \textit{tokensReceived}       & 0x0023de29                                  \\ \hline
EIP-1820          & \textit{canImplementInterfaceForAddress}   & 0x249cb3fa                             \\ \hline
\end{tabular}}
\label{tab:eip}
\end{table}


\subsection{Cross-contract Static Analysis}
In this subsection, we describe how we use decompiled intermediate representation (IR) and intra-procedure information to construct the xCFG and xCG of the attacker contract and its interacting contracts. Algorithm 1 presents an overview of how to identify the attacker contract. We first construct the call chain of every public function that contains an external call $E_f$ using the extracted flow information (L1). Then, we use the depth-first search (DFS) algorithm for each public function $f$ to find its external call target contracts and functions to construct the xCFG and xCG (L2-L3). We apply the tainted source identification rules to find the tainted source $s$ in the input contract's function (L4). Then, we use sink identification rules for each call chain in the constructed xCG to find sensitive variables $t$ that can cause the attack (L6), such as function arguments flowing to callee variables. To determine whether contract \textit{C} can successfully make the other contract call back to itself, we use the transfer rules from source to sink to obtain reachability (L7) and find possible attack call chains.

\begin{table}[htbp]\small\addtolength{\tabcolsep}{-3pt}
	\normalsize
\vspace{0.1in}
\centering
    \scalebox{0.9}{
	\begin{tabular}{ l }
		\hline\noalign{\smallskip}
		\textbf{Algorithm 1:} Cross-contract Static Analysis for Identifying Attacks\\
		\noalign{\smallskip}\hline\noalign{\smallskip}
        \textbf{input}: $C$, the input contract bytecode \\
        \textbf{output}: $AC \leftarrow []$, the list of possible attack call chains \\
        1: $E_f \leftarrow findFunctionsWithExternalCall(C)$  \\
        2: \textbf{for each} function $f \in E_f$ \textbf{do} \\   
	3: \ \ \ \   $P_c \leftarrow searchCallPathsByDFS(f)$ \\
        4: \ \ \ \   $s \leftarrow getSource(f)$ \\
	5: \ \ \ \  \textbf{for each} path $p \in P_{c}$ \textbf{do}\\
	6: \ \ \ \ \ \ \ \ $t \leftarrow getPossibleSink(p)$ \\ 
        7: \ \ \ \ \ \ \ \ \textbf{if} $isReachable(s,t,rules)$ \\ 
	8: \ \ \ \ \ \ \ \ \ \ \ \ $AC \leftarrow AC \cup p, break$ \\
        9: \textbf{return} $AC$ \\
	\hline
	\end{tabular} }\label{alg:algorithm1}
\end{table}

\subsubsection{xCFG Construction \& Call Chain Gathering}
To perform static dataflow analysis, we first need to construct the xCG and xCFG of the input contract. We obtain the bytecode of the interacted contracts based on the contract address identified during decompilation. Then, for every external call of the input contract, we find the call-target contract address and function signature. Based on the tuple <\textit{callsite}, \textit{caller\_address}, \textit{caller\_funcSign}, \textit{target\_contract}, \textit{target\_funcSign}>, we find every call site that executes the \textit{CALL} opcode and its call target. We then use the DFS algorithm to construct the xCFG and gather all possible call chains originating from the public functions of the input contract to construct xCG for the dataflow analysis.

\subsubsection{Cross-contract Dataflow Analysis}\label{sec:cross-contract-analysis}
We define the tainted source, sink site, and transfer rules in our dataflow analysis for identifying attacks in Table~\ref{tab:rule}. For a contract \textit{C} and its external call \textit{f} with arguments set \textit{A} in the example, we define all external call arguments $A_s$ in contract \textit{C} as tainted. Then, for every contract called, we mark every callee variable that determines the call target address $C_t$ as the sink site. We apply the rules shown in Table~\ref{tab:flow} to determine whether the tainted source can flow to the sink site. If there is a possible path for that dataflow path, it is possible that the input contract can make the called contract call a specific address that the attacker designed, which helps us find whether there is reentrancy.

\begin{table}[htbp]
\centering
\caption{Dataflow Rules for Identifying Attacker Contracts}

\begin{tabular}{l||m{6.5cm}}
\hline
\textbf{Example}               & \textit{C.f(A)}, \textit{f} is an external call to contract \textit{C}; \textit{A} is the set of arguments \\ \hline
\textbf{Source}         & external call arguments set $A_s$ of input contract   \\ \hline
\textbf{Rules} & 5 intra-procedure dataflow rules in Table~\ref{tab:flow}                                    \\ \hline
\textbf{Sink} & callee $C_t$ of external calls in called contracts                  \\ \hline
\end{tabular}\label{tab:rule}
\end{table}

\subsection{Attacker Contract Detection}
In this subsection, we present specific rules for detecting reentrancy attacks in the attacker contract using flow information and cross-contract dataflow analysis.

The reentrancy attack can be reflected in the call chain that we recover by cross-contract static analysis. Specifically, we detect the attacker contract that can perform a reentrancy attack in three steps.
Step 1: We first determine whether there is a call path that causes tainted variables to flow to the sink site using the rules we designed in the cross-contract analysis illustrated in Section~\ref{sec:cross-contract-analysis}. For example, in Figure~\ref{fig:flow}, the attacker can call function \textit{bar()} (L2-L10) to invoke external call \textit{target.foo()} (L16-L22). The function argument \textit{v1} (L16) can flow to the callee of the external call \textit{v1.hook(v2)} (L19) in contract \textit{target}, which means contract \textit{from} can manipulate the call target of \textit{v1.hook()}. We then find the target call function \textit{hook()} (L11-L12) of the reachable sink site \textit{v1.hook()} (L19).
Step 2: We determine whether there is an implementation of the function \textit{hook()} (L11-L12) of the input contract, i.e., contract \textit{from}.
Step 3: If Step 2 is true, we find the call target contract address and function signature to determine whether they are visited in the call path, and judge whether there is a reentrancy attack. In the example, contract \textit{from} calls back to the function \textit{foo()} again in the function \textit{hook()} to perform reentrancy.

We use the following expression to define the conditions for a reentrancy attack. We first find the reachability of the sink site in contract \textit{tar}, which calls the function \textit{f} of contract \textit{to}. The address of contract \textit{to} can be set by the attacker, which is the tainted source denoted by $A_s$. Furthermore, the function \textit{f} should be implemented in the attacker contract's public function list $C_F$, and the external calls $f_{EC}$, including the call target address and function signature, should be visited in the call path. If all conditions are met, we consider it a reentrancy attack.

$$Reentrancy \Leftarrow  \dfrac{Reachable(tar,f,to), \; to \in A_s}{f \in C_F, \; f_{EC} \in Visited}$$

\section{Evaluation}\label{sec:evaluation}
In this section, we evaluate the effectiveness of BlockWatchdog based on the ground truth dataset collected from attack reports and a large-scale dataset obtained through blockchain transaction replay.

\subsection{Evaluation Setup}
The experiment was conducted on a server running Ubuntu 20.04.1 LTS and equipped with 18 Intel(R) Core(TM) i9-10980XE CPUs @3.00GHz and 250 GB memory.

\para{Dataset} We use two datasets to evaluate BlockWatchdog. \textit{The first} one is the ground truth dataset, which comprises reentrancy attacker contracts that we collect by analyzing the attack incidents reports. This dataset contains 18 attacker contracts from 15 incidents. \textit{The second} large-scale consists of 421,889 real-world contract bytecode (both creation and runtime). These contracts are obtained via replaying transactions from block number 10 million to 15.5 million on the Ethereum mainnet. 

\para{Evaluation Metrics} We summarize the following research questions (RQs) to evaluate BlockWatchdog.
\begin{enumerate}[RQ1.]
	\item How effectively is BlockWatchdog in detecting reentrancy attacks in the ground-truth dataset?
        \item How is the performance of BlockWatchdog in finding attacker contracts in the large-scale dataset?
	\item How much financial loss is caused by the identified attacks, and are vulnerability detection tools able to find those vulnerable victims?
\end{enumerate}

\subsection{Answer to RQ1: Effectiveness on the Ground Truth Dataset}
To answer RQ1, we run BlockWatchdog on our ground truth dataset with 18 attacker contracts from our collected reports. BlockWatchdog correctly reports 15 out of 18 reentrancy attacker contracts, as shown in Table~\ref{tab:dataset1}. 
The second and third columns show the name of the DApp and the time it was attacked; the fourth and fifth columns represent the loss of the attack and the platform the DApp was deployed.
Columns sixth to eighth show the address of the exploiter, the attacker contract, and the victim contract in the attack, respectively. The last column denotes whether BlockWatchdog can identify the attacker contract.
It is noticeable that we find two exploiters (0xce and 0x80) in the Cream Finance attack from our collected incidents, and these two exploiters deployed 2 (0xbd and 0x38) and 1 (0x32) attacker contracts, respectively, to attack the same victim contract (0xce). 
Similarly, the exploiter (0x61) that attacked the Fei protocol and Rari DApps deployed two attacker contracts (0xE3 and 0x32) to attack the victim contract (0xfb).
We find that BlockWatchdog fails to identify three reentrancy attacker contracts due to two reasons.
The first factor is the inability of BlockWatchdog to recover the call chain when the call target addresses and functions in the attacker contracts are obtained from memory. Given that the memory value cannot be determined via static analysis, BlockWatchdog relies on the constant value of callee addresses or call target function signatures for complete call chain recovery.
In the cases of the Fei Protocol and Rari incident, BlockWatchdog cannot identify them because the call target contract address is loaded from the memory.
However, the memory value can only be determined during runtime, making BlockWatchdog fail to deduce the call target and recover the call chain.
The second reason pertains to the limitations in the function signature identification of BlockWatchdog. Since BlockWatchdog is based on Elipmoc~\cite{grech2022elipmoc}, which may not be able to recover all function signatures, this limitation may lead BlockWatchdog to fail to identify some external calls. For example, BlockWatchdog cannot detect the reentrancy attack in the Spankchain incident, as the call target function \textit{LCOpenTimeout()} is not identified in the victim contract. As the dataflow procedure ends with an unknown target, BlockWatchdog fails to recover the call chain and identify the reentrancy.
Despite such cases, the recall of BlockWatchdog reaches 83.3\% in our experiments.

\begin{table*}[!htbp]
\centering
\caption{Attacker Contracts in the Dataset and Detection Results of BlockWatchdog (BW)}\label{tab:dataset1}
\begin{tabular}{llllllllc}
\toprule
\textbf{\#} & \textbf{DApp} & \textbf{Attack Time} & \textbf{Loss (\$)} & \textbf{Platform} & \textbf{Exploiter} & \textbf{Attacker Contract} & \textbf{Victim Contract} & \textbf{BW} \\
\cmidrule(r){1-2} \cmidrule(r){3-5} \cmidrule(r){6-8} \cmidrule(r){9-9}
1  & Spankchain & 2018/10/9   & 38K       & ETH      & 0xcf267eA3f1eb & 0xc5918a927C4F    & 0xf91546835f75  & {$\times$} \\
\rowcolor{gray!30} 2  & Uniswap  & 2020/4/18   & 220K     & ETH      & 
0x60f3FdB85B2F & 0xBD2250D713bf    & 0x1f9840a85d5a  & \checkmark \\
3  & Lendf.Me  & 2020/4/19   & 24.7M     & ETH      & 0xa9bf70a420d3 & 0x538359785a8D    & 0x0eEe3E3828A4  & \checkmark \\
\rowcolor{gray!30}4  & Akropolis & 2020/11/12  & 2M        & ETH      & 0xe2307837524D & 0x38c40427efbA    & 0x1cec0e358f88  & \checkmark \\
5  & DeFiPie   & 2021/7/12   & 350K      & BSC      & 0xf6f43f77ef9e & 0x6d741523F1Fc    & 0x607C794cDa77  & \checkmark \\
\rowcolor{gray!30}6  & xSurge      & 2021/8/16   & 25M        & BSC    & 0x59c686272e6f & 0x1514aaa4dcf5     & 0xE1E1Aa58983F  & \checkmark \\
\multirow{3}{*}{7} & \multirow{3}{*}{Cream Finance} & \multirow{3}{*}{2021/8/31} & \multirow{3}{*}{5M} & \multirow{3}{*}{ETH} & \multirow{2}{*}{0xce1f4b4f1722} & 0xbd51Cb8c06F7 & \multirow{3}{*}{0xce1f4b4f1722} & \checkmark \\
 &  &  &  &  &  & \multicolumn{1}{l}{0x38c40427efbA} &  & \multicolumn{1}{c}{\checkmark} \\
 &  &  &  &  & \multicolumn{1}{l}{0x8036EbD0Fc9C} & \multicolumn{1}{l}{0x32d77947aACa} &  & \multicolumn{1}{c}{\checkmark} \\
\rowcolor{gray!30}8  & Grim Finance      & 2021/10/16   & 30M        & FTM    & 0xDefC385D7038 & 0xb08cCb39741d     & 0x279b2c897737  & \checkmark \\
9  & Visor Finance      & 2021/12/21   & 8.2M        & ETH    & 0x8efab89b497b & 0x10C509AA9ab2     & 0xc9f27a50f825  & \checkmark \\
\rowcolor{gray!30}10  & Paraluni      & 2022/3/13   & 1.7M        & BSC    & 0xA386F30853A7 & 0x4770b5cb9d51     & 0x94bC1d555E63  & \checkmark \\
11  & Agave \& Hundred      & 2022/3/15   & 5.5M        & ETH    & 0xcE1F4B4F1722 & 0x38c40427efbA     & 0xf8D1677c8a0c  & \checkmark \\
\rowcolor{gray!30}12 & Revest      & 2022/3/27   & 120K        & ETH    & 0xef967ece5322 & 0xb480ac726528     & 0x2320a28f5233  & \checkmark \\
\multirow{2}{*}{13} & \multirow{2}{*}{Fei Protocal \& Rari} & \multirow{2}{*}{2022/4/30} & \multirow{2}{*}{80.3M} & \multirow{2}{*}{ETH} & \multicolumn{1}{l}{\multirow{2}{*}{0x6162759eDAd7}} & \multicolumn{1}{l}{0xE39f3C40966D} & \multicolumn{1}{l}{\multirow{2}{*}{0xfbD8Aaf46Ab3}} & \multicolumn{1}{c}{$\times$} \\
 &  &  &  &  & \multicolumn{1}{l}{} & \multicolumn{1}{l}{0x32075bAd9050} & \multicolumn{1}{l}{} & \multicolumn{1}{c}{$\times$} \\
\rowcolor{gray!30}14 & Omni      & 2022/7/10   & 1.4M        & ETH    & 0x00000000c251 & 0x3c10e78343c4     & 0x3c10e78343c4  & \checkmark \\
15 & SushiBar      & 2022/10/25   & 15K        & ETH    & 0x8ca72f46056d & 0x9C5A2A643152     & 0x2321537fd8EF  & \checkmark \\
\bottomrule
\end{tabular}
\end{table*}

\subsection{Answer to RQ2: Attacker Contracts Detection in a Large-scale Dataset}
To answer RQ2, we ran BlockWatchdog on 421,889 smart contracts obtained from blockchain transaction replay. The experimental results show that BlockWatchdog reports 253 contracts as attacker contracts.
We next evaluate the performance of BlockWatchdog, using a manual labeling process. Two authors manually inspect the 253 samples reported as attacker contracts, following a four-step procedure. \textit{Firstly}, the decompiled intermediate representation (IR) of the attacker contract was thoroughly examined. \textit{Secondly}, the function that performs the attack reported by BlockWatchdog is located. \textit{Thirdly}, the call chains reported by BlockWatchdog are checked. \textit{Finally}, based on these examinations, it is determined whether a contract is an attacker contract, and the associated financial loss is recorded.
The labeled results show that 113 samples are indeed attacker contracts that perform reentrancy attacks. 

In the large-scale experiment, some metrics are collected to help us evaluate BlockWatchdog more comprehensively. For the structure of the constructed xCFG and xCG, the average number of visited contracts and the call depth of the recovered xCG are 0.95 and 0.21, respectively. The same indicators of true attacker contracts are 7.66 and 2.55, respectively.
This indicates that many attacker contracts directly hardcode the victim contracts' addresses for implementing the attack.
In addition, the average detection time for BlockWatchdog to analyze an attacker contract is 17.66 seconds. 
An attacker contract can interact with a maximum number of 105 contracts in a single call chain, with the maximum call depth reaching 21.
Such deep call chains involving multiple contracts and functions expose the limitations of simple pattern-based rules in covering complex reentrancy attacks.


\begin{table}[!htbp]
\centering
\caption{Top 5 Hook Functions and Call Target Functions in Identified Attacker Contracts with Occurrence Times}
\resizebox{\columnwidth}{!}{
\begin{tabular}{l|c||l|c}
\hline
\textbf{Hook Function }         & \textbf{Times} & \textbf{Call Target} & \textbf{Times} \\ \hline
\textit{uniswapV2Call}       & 73    & \textit{balanceOf}             & 111    \\ \hline
\textit{onFlashLoan}           & 13     & \textit{transfer}              & 102    \\ \hline
\textit{onERC721Received}         & 9     & \textit{approve}             & 53    \\ \hline
\textit{delegatedTransferERC20}      & 2     & \textit{withdraw}      & 25    \\ \hline
\textit{onERC1155Received} & 2     & \textit{deposit}             & 23   \\ \hline
\end{tabular}} \label{tab:hookcall}
\end{table}

Table~\ref{tab:hookcall} shows the signatures of the call targets in the hook function of the attacker contracts from our labeled dataset. The \textit{uniswapV2Call()} is a required hook function in Uniswap V2~\cite{uniswapv2}, while \textit{onFlashloan()} is declared by ERC3156~\cite{eip3156}.
We also have some interesting findings related to the design of attacker contracts. For example, 19 attacker contracts were designed with attack logic to hack victim contracts, but these attempts failed, resulting in reverted transactions. Most of these contracts only have two transactions, i.e., the contract creation transaction and the failed attack transaction. 
Moreover, some exploitation function names, such as \textit{Attack}, \textit{Rugpull}, \textit{Exploit}, and \textit{Trigger}, are commonly used in attacker contracts. This interesting discovery provides information for function signature or name identification to find potential attacker contracts, and provides insights into the development preferences of hackers.

\para{False positives} We identify two types of false positives generated by BlockWatchdog when detecting reentrancy attacker contracts. The first type involves the use of \textit{getter} functions to make external calls in the reentrant hook function, without performing any other profitable external call operations. The second type involves the usage of a permission check mechanism, where some contracts use \textit{msg.sender} as the transfer target or to constrain the caller of the hook function, with no intention of making external calls to attack others. BlockWatchdog reports all cases based on the reentrancy path to minimize false negatives, which may generate false positives.

\vspace{-0.1cm}
\subsection{Answer to RQ3: Financial Loss of Victims}
During the labeling process, we found that the total financial loss caused by the true positive attacks was 908.4 million USD. This loss comprised approximately 840 Ethers (about 1.7 million USD) and tokens worth 906.9 million USD. Notably, the loss of tokens accounted for 99.8\% of the total financial loss, indicating that new types of reentrancy attacks are primarily caused by poor designs when using and transferring ERC tokens, rather than Ether transfers.

Table~\ref{tab:loss} shows the top 5 attacks identified by BlockWatchdog, ranked by their financial loss. We find that vulnerable contracts may be attacked multiple times if they have been exploited successfully once.
Specifically, BlockWatchdog identified 9, 4, and 3 attacker contracts that were deployed to attack projects Cream Finance~\cite{cream}, Omni~\cite{omni_hacker}, and Visor Finance~\cite{ivisor}, respectively, resulting in a total financial loss of 906.7 million USD. Cream Finance was hacked by the hook function \textit{tokensReceived()} in the attacker contract. It implements multiple token-borrow logic to perform the flash loan attack in a reentrancy call path, which demonstrates the complexity of the new types of reentrancy attacks.
Omni was attacked by attacker contracts that implemented reentrancy logic in the hook function \textit{onERC721Received()} when transferring NFTs. Hackers repeatedly minted, borrowed, and withdrew NFTs before changing the liquidation state.
Visor Finance was exploited by attacker contracts that implemented \textit{delegatedTransferERC20()}, which is a user-defined hook function, as shown in Figure~\ref{fig:ivisor}. In addition, BlockWatchdog identified 40 zero-day attacker contracts and a total of 159 victim contracts, which were targeted by the attacker contracts to perform reentrancy attacks.

\begin{table}[htbp]
\setlength{\abovecaptionskip}{0.2cm}
\centering
\caption{Top 5 Attacks Ranked by Financial Loss}
\resizebox{\columnwidth}{!}{
\begin{tabular}{l||l|l}
\hline
\textbf{Attaker Contract}                               & \textbf{Hook Function}              & \textbf{Loss (USD)}  \\ \hline
0x10c509aa9ab2      & \textit{delegatedTransferERC20} & 904 million     \\ \hline
0xc51bdc9aebba
     & \textit{tokensReceived}         & 0.56 million           \\ \hline
0x86f28c7030bd
     & \textit{onERC721Received}       & 0.28 million           \\ \hline
0xbc82ab5a8223
     & \textit{tokensReceived}         & 0.28 million            \\ \hline
0x3292818dB514
 & \textit{uniswapV2Call}          & 0.28 million        \\ \hline
\end{tabular}}
\label{tab:loss}
\end{table}

\section{Discussion}\label{sec:discussion}
In this section, we first give a case study to show how BlockWatchdog can enhance the security of Ethereum by identifying attacker contracts. 
We then present the capability of BlockWatchdog to find vulnerabilities and discuss the limitations of our work.

\subsection{Case Study}\label{sec:case} 

Figure~\ref{fig:revest} shows a code snippet of the attacker contract that aimed to attack Revest~\cite{revest} identified by BlockWatchdog. This attacker contract was first deployed at Mar-27-2022 01:10:05 AM +UTC, and about half an hour later, the first attack was launched at Mar-27-2022 01:41:46 AM +UTC. 
Specifically, the function with signature \textit{0xdd869c35} was invoked by the exploiter to call victims. 
Then, the call chain was turned back and controlled by the attacker contract twice through function invocation by the hook functions, i.e., \textit{uniswapV2Call()} (L1-L4) and \textit{onERC1155Received()} (L6-L17).
The attacker contract queried for the next NFT id (function \textit{getNextId()} (L12)) and deposited another one via the function \textit{depositAdditionalFNFT()} (L13).
Finally, another NFT was successfully minted to the attacker without being paid in this reentrancy attack.
It takes BlockWatchdog 64 seconds to identify this attacker contract and its victim contract.
The time gap of half an hour between the deployment of the attacker contract and the actual attack far exceeds the time costs of the detection process. This provides an opportunity for whitehats to front-run the hackers and protect the vulnerable deployed contracts. Specifically, whitehats can copy the bytecode of the attacker contract to perform the imitation attack~\cite{qin2023blockchain} before the real attack is launched. Therefore, the intended profits of hackers are extracted by whitehats, thus saving possible financial losses.
In addition, BlockWatchdog takes 25 seconds to detect this attacker contract (shown in Figure~\ref{fig:ivisor}) and the potentially exploitable victim in Visor Finance~\cite{ivisor}. Since the attack occurred about 6 minutes after the deployment of the attacker contract, the protection can be performed in this time gap.
\begin{figure}[htbp]
\begin{lstlisting}
 function uniswapV2Call(address varg0, uint256 varg1, uint256 varg2, bytes varg3) public nonPayable { 
   v18, v19 = stor_18_0_19.call(0x2e236bc, address(this), 1, ...);
   ...
   v39 = stor_18_0_19.withdrawFNFT(v36, 1 + v2[0]);}
 
 function onERC1155Received(address varg0, address varg1, uint256 varg2, uint256 varg3, bytes varg4) public nonPayable { 
   ...
   if (_onERC1155Received != 0) {
     if (_onERC1155Received == 1) {
       v0 = 0x2007(_onERC1155Received);
       _onERC1155Received = v0;
       v1, v2 = stor_19_0_19.getNextId();
       v3, v4 = stor_18_0_19.depositAdditionalToFNFT(v2 - 1, stor_4, 1);}
    } else {
       v5 = 0x2007(_onERC1155Received);
       _onERC1155Received = v5;}
    return 0xf23a6e61;}
\end{lstlisting}

\caption{The attacker contract that hacked the Revest.}
\label{fig:revest}
\end{figure}

The above case study shows the practicality of timely protection by using our BlockWatchdog. In the blockchain system, attacks are irreversible, making it critical to detect potentially threatening contracts before the start of any transaction. BlockWatchdog can quickly identify such vulnerable contracts, thereby improving the security of the Ethereum ecosystem in two ways.

Firstly, our tool can be used for the real-time detection of attacker contracts on Ethereum, allowing security firms to report suspicious attacker contracts in minutes. Since we do not require transaction information, it is possible to prevent attacks.
Whitehats~\cite{whitehat}, Etherscan~\cite{etherscan}, and security analysis firms, such as Consensys~\cite{consensys}, can quickly identify potential attacker contracts and victims with attack footprints reported by BlockWatchdog (17.66 seconds on average) before attack transactions are sent (as shown in the above case).
However, it is crucial to address the ethical implications of such preemptive actions in this paper. While our approach facilitates early detection and provides an opportunity for rapid response, we consciously do not perform on-chain frontrunning that could be deemed as ethical issues or potential disruptiveness to the blockchain's integrity. Furthermore, we use transaction replay only to retrieve the bytecode of the deployed smart contracts.

Secondly, our tool can find new types of reentrancy vulnerabilities that other tools may have missed. Security firms can use BlockWatchdog to identify undiscovered attacker contracts and reentrancy vulnerabilities in practice. Platforms like Etherscan label the attacker's EOA as an attacker account, and warn about newly deployed contracts and related transactions.

For contract developers, to prevent their contracts from being attacked after the deployment, the external call target contract's address should be verified in ``sensitive'' functions, e.g., token transfer or swap logic, thus blocking the premeditated attacks that malicious attacker contracts could launch.

\subsection{Capability of Finding Vulnerabilities}
BlockWatchdog finds a significant number of exploitable victim contracts with exploitable vulnerabilities, and most of them cannot be detected by current tools.
We choose seven reentrancy detection tools to detect victim contracts: Mythril~\cite{mythril}, NFTGuard~\cite{nftdefects}, Oyente~\cite{luu2016making}, Sailfish~\cite{rao2012sailfish}, Securify1~\cite{tsankov2018securify}, Securify2~\cite{securify2}, and Smartian~\cite{choi2021smartian}, referring to the four select rules~\cite{zheng2023turn,nftdefects}, i.e., (1) availability of the tool source code; (2) usability of the command-line interface for large-scale experiment; (3) supporting Solidity source code; (4) ability to report vulnerable code location for manual examination.
We run these seven tools on 159 victim contracts, and all outputs are given in our open repository.

The results show that only Mythril and Sailfish report 1 and 17 victims, respectively, while the other five tools do not report any reentrancy vulnerability. Therefore, only 18 out of 159 (11.3\%) victims can be detected by current tools.
As BlockWatchdog extracts more features from attacker contracts, which may be missed by other works, BlockWatchdog can more effectively identify exploitable contracts with reentrancy vulnerabilities in practice with better performance and generalizability.
There are three reasons for the results. 
First, many identified attacker contracts extract tokens but not Ethers from victims. However, most existing tools focus on \textit{call.value()}, which only involves Ether transfer. Second, rule-based methods struggle to cover issues caused by patterns that have not been previously reported, and it is even harder to identify reentrancies related to user-defined interfaces that cannot be generalized into a detection pattern. Third, existing tools detect reentrancy based on state modification inconsistency. However, new reentrancy attacks, \textit{e.g}., read-only reentrancy, can make use of just view functions, e.g., \textit{balanceOf()} and \textit{getPrice()} to implement reentrancy.
These view functions do not modify any state and are usually not protected, making existing tools difficult to detect.



\subsection{Limitations}
Despite the strengths of BlockWatchdog, we identify three potential limitations. \textit{First}, BlockWatchdog reports whether an input contract is an attacker contract or not based solely on static analysis without transaction information. We can recover the possible call chains of the input contract, including transaction data, which can help enhance the precision of detection. As we aim to deploy BlockWatchdog as a real-time detection platform for Ethereum in the future, transactions are considered as additional information to achieve a more precise detection result
\textit{Second}, we identify interacted contracts from the constant address or storage, and by default, we rely on the assumption that the attacker contract hardcodes the victims' addresses. Although we find that all the collected cases belong to these two scenarios, it is possible that the attacker contracts can set the target contract in the arguments of the function, which we cannot obtain through static analysis, thus leading to false negatives.
In addition, our taint analysis does not account for conditional checks when tracing paths, which yields false positives.
\textit{Third}, we summarize the attack types and design the detection rules based on the reported attack incidents. Therefore, it is possible that we may not cover new attack types that have not yet occurred or been reported. 
Overall, despite the above limitations, we focus on detecting new types of reentrancy vulnerabilities that other tools cannot cover. Other types of attacks that involve attacker contracts mentioned in Section~\ref{sec:reen} are not addressed in this paper, which we will cover in future work.

\subsection{Threats to Validity}
Regarding our experiment, the manual labeling process may have introduced errors in differentiating false positives and true positives. However, we have used a double-check process to mitigate this issue and updated the labeled dataset in a timely manner to ensure accuracy. We validated whether there was an attack on Etherscan using the call chains reported by BlockWatchdog, collaborated with the transaction trace recorded on the online transaction trace explorer Phalcon~\cite{phalcon}, and recorded financial loss according to the transaction information obtained from Etherscan.
Another threat to validity is that we did not verify that vulnerable contracts can be exploited due to ethical concerns about attacking them. We intend to address this in our future work.

There are some other types of attacker contracts, e.g., Bad Randomness and Flashloan, as we mention in~\ref{sec:attackercontracts}, which are not covered by this work.
The design of our BlockWatchdog focuses on call flow analysis, which makes it possible to be extended to detect Flashloan attacker contracts that also contain some flow features.
However, more contract semantics and operational features should be analyzed to cover other attacker contract types like Bad Randomness, which requires understanding specific contract behaviors.


Despite the limitations mentioned above, BlockWatchdog has detected 40 zero-day attacker smart contracts that were previously unreported, which had reentrancy features. Since BlockWatchdog does not require transaction information and can provide detection results within minutes, it can monitor newly deployed contracts and detect attacker contracts before they can execute attacks, preventing potential attacks and heavy financial losses.

\section{Related Work}\label{sec:rw}
\para{Reentrancy detection tools for smart contracts}
As reentrancy is one of the notorious vulnerabilities in smart contracts~\cite{rodler2018sereum}, many program analysis tools have been developed to detect such issues by static analysis or dynamic testing~\cite{bertolino2007software,6963470,cadar2008klee}. The goal of these tools is to prevent vulnerable contracts from being deployed on the blockchain.
For example, Oyente~\cite{luu2016making}, Securify~\cite{tsankov2018securify}, Mythril~\cite{mythril}, and Sailfish~\cite{rao2012sailfish} use
static analysis technologies to discover reentrancy vulnerability. 
In addition, there are dynamic testing and analysis tools such as ContractFuzzer~\cite{jiang2018contractfuzzer}, sFuzz~\cite{nguyen2020sfuzz}, Smartian~\cite{choi2021smartian}, RLF~\cite{su2022effectively}, and ReGuard~\cite{liu2018reguard}, and approaches based on machine learning like ReVulDL~\cite{zhang2022reentrancy}.
As contract code and vulnerabilities become more complex, there are works that focus on cross-contract analysis, such as Clairvoyance~\cite{xue2020cross} and SmartDagger~\cite{liao2022smartdagger}.
However, many of these tools suffer from high false positive rates and may not identify real vulnerable contracts in practice~\cite{perez2021smart}.

To our knowledge, BlockWatchdog is the first detection tool to identify attacker contracts and their call chains. This feature enables the tool to detect real vulnerable victim contracts in practice. Additionally, BlockWatchdog addresses the limitation of current reentrancy detection capability~\cite{zheng2023turn}, and can detect complex reentrancy attacks in the wild.

\para{Security analysis on attack incidents on Ethereum}
Prior research on attack analysis on Ethereum has focused on understanding attack incidents at the transaction level. For example, Torres et al.~\cite{torres2021frontrunner} conducted an empirical study of front-running attacks on Ethereum, while Zhou et al.~\cite{zhou2020ever} evaluated real-world attacks and defenses in the Ethereum ecosystem. They analyzed how attackers have destroyed applications in Ethereum and discussed how to defend against attacks from the victim contract's perspective. Su et al.~\cite{su2021evil} focused on DApp security and analyzed related transactions to understand how to detect attacks through transaction analysis. They developed a tool called DEFIER that can identify the stage of a potential attack. 
However, BlockWatchdog can detect attacker contracts without any attack transactions, making it possible to prevent financial loss.

\section{Conclusion and Future Works}\label{sec:conclusion}
In this paper, we present BlockWatchdog, a tool for detecting reentrancy attacker contracts and identifying vulnerable victims with reentrancy vulnerabilities. To reduce false positives, we use the detection of attacker contracts as an entry point, and identify vulnerable victim contracts based on callback flow. To design BlockWatchdog, we conducted an empirical study to understand the attack logic used by hackers in attacker contracts.
BlockWatchdog disassembles a contract's bytecode and monitors all potential call chains that initiate from its public functions and extend to accessible contracts and functions. 
Besides, BlockWatchdog formulates the xCFG and xCG to facilitate cross-contract data flow analysis between different procedures and to determine whether the callback flow can be exploited by malicious contracts to execute a successful reentrancy attack.
Our experiment results demonstrate that BlockWatchdog effectively detects 113 attacker contracts among 421,889 real-world contracts and identifies 159 victim contracts with reentrancy vulnerabilities. These vulnerable contracts contain Ethers and tokens worth approximately 908.6 million USD. Only 18 of them are identified by other detection tools.

Furthermore, we reveal all potential call chains of the 421,889 real-world contracts, and whether they contain external calls in hook functions identified by BlockWatchdog. The detection results can be helpful for further security analysis. In the future, we plan to deploy BlockWatchdog for real-time detection purposes, in order to find more vulnerable contracts in practice and help prevent them from being attacked. In addition, we will extend BlockWatchdog to cover new types of attacks, remaining effective in the face of ever-evolving threats on Ethereum.

\begin{acks}
This research/project is supported by the National Key R\&D Program of China (2022YFB2702203), the National Natural Science Foundation of China (No. 62302534 and No. 62332004), and the Ant Group Research Fund.
\end{acks}

\bibliographystyle{ACM-Reference-Format}
\bibliography{ref}










\end{document}